\documentclass[12pt,aps,prb,preprint]{revtex4}   
\usepackage{amsmath}    % need for subequations
\usepackage{graphicx}   % for figures
\begin{document}
\title{An ab initio derivation of the electromagnetic fields of a point charge in arbitrary motion}
\author{Ashok K. Singal}
\affiliation{Astronomy and Astrophysics Division, Physical Research Laboratory,
Navrangpura, Ahmedabad - 380 009, India }
\email{asingal@prl.res.in}
\date{\today}
\begin{abstract}
Electromagnetic fields of an accelerated charge are derived from the first principles using Coulomb's law and the 
relativistic transformations. The electric and magnetic fields are derived first for an instantaneous rest frame of the 
accelerated charge, without making explicit use of Gauss's law, an approach different from that available in the literature. 
Thereafter we calculate the electromagnetic fields for an accelerated charge having a non-relativistic motion. 
The expressions for these fields, supposedly accurate only to a first order in velocity $\beta$, 
surprisingly yield all terms exactly for the acceleration fields, only missing a factor $1-\beta^2$ in the velocity fields. 
The derivation explicitly shows the genesis of various terms in the field expressions, 
when expressed with respect to the time retarded position of the charge. A straightforward transformation 
from the instantaneous rest frame, using relativistic Doppler factors, yields expressions of the electromagnetic fields for the 
charge moving with an arbitrary velocity. The field expressions are derived without using Li\'{e}nard-Wiechert 
potentials, thereby avoiding evaluation of any spatial or temporal derivatives of these potentials at the retarded time.
\end{abstract}
\maketitle
\section{Introduction}
The electromagnetic (EM) fields of a moving charge are formally calculated from Li\'{e}nard-Wiechert potentials.\cite{1,2,3,12} 
The mathematics, quite tricky and involved since derivatives of the potentials need to be evaluated at the retarded 
time, can be quite a deterrent to the initiate. 
The expression for the fields in the instantaneous rest frame of an accelerated charge can be worked out from physical 
arguments.\cite{9,4,5,8} It is generally believed that a relativistic transformation of fields from the instantaneous rest frame 
to an inertial frame in which charge has an arbitrary velocity could be quite laborious.\cite{7} 
EM fields for the special case of acceleration parallel to the 
velocity vector have been derived\cite{10,11} without using Li\'{e}nard-Wiechert potentials.  Huang and Lu \cite{6} attempted 
the more general case but ended up with wrong expressions for the 
EM fields though they claimed to be giving ``the exact expression'' for radiation of an accelerated charge. To bypass the 
mathematical cumbrousness, Padmanabhan\cite{7} used an alternate approach where the expressions for EM field were derived 
in an indirect manner, by finding a general covariant 4-vector function of position, velocity and 
acceleration of the charge which coincided with the EM field values in the instantaneous rest frame. The approach though elegant 
is not immediately obvious. It may thus be still worthwhile to have EM field expressions transformed directly from the rest-frame 
values using a 3-vector language, which should be transparent to the reader. 

We show here that the standard text-book expressions for the EM fields of an accelerated charge can be derived in 
a fairly easy and straightforward manner, without using Li\'{e}nard-Wiechert potentials, thereby avoiding evaluation of any 
spatial or temporal derivatives of these potentials at the retarded time. We start with the radial Coulomb field of a 
stationary charge and then making use of the relativistic transformations, in particular that of EM fields, we derive the fields 
for an  accelerated charge in its instantaneous rest frame where transverse field components proportional to acceleration show up. 
The presence of such an electric field component proportional to acceleration was shown by Thomson,\cite{9}  
using a physical picture in the pre-relativity days, employing the concept of electric field lines representing Faraday 
(flux) tubes. A modern derivation using Gauss's law is now available in many  
text-books.\cite{4,5,8} We shall derive the transverse components for both electric and magnetic fields, in the same  
spirit but from a different perspective, without explicitly using Gauss's law. Thereafter we get field expressions for a 
slowly moving charge using non-relativistic transformations and surprisingly the expression for the acceleration fields turn 
out to be exactly the same as for a relativistically moving charge, which we derive rigorously in a later section.
We have used Gaussian system of units throughout.
\section{Electromagnetic fields of an accelerated charge in an instantaneous rest-frame}
\subsection{A uniformly moving charge}
The electric field of a stationary charge (Coulomb's law) in an inertial frame, say $\cal S'$, is,
\begin{eqnarray}
{\bf E'} & = & \frac{e}{R'^{2}} {\bf n'}
\end{eqnarray}
where ${\bf n}'$ is a unit vector in radial direction from the position of the charge.

The field expressions for a charge moving with a uniform 
velocity ${\bf V}={\mbox{\boldmath $\beta$}}c$  can be derived by a relativistic transformations of the fields to a frame $\cal S$, 
with respect to which $\cal S'$ has a uniform velocity ${\bf V}$. 
From the Lorentz transformation of EM fields,\cite{1,2,3,4} we have 
\begin{eqnarray}
\nonumber 
{\bf E} = {\bf E}'_\parallel + \gamma[{\bf E}'_\perp - {\mbox{\boldmath $\beta$}}\times\bf B'], 
& \,\,\,{\bf B} =  {\bf B}'_\parallel + \gamma[{\bf B}'_\perp + {\mbox{\boldmath $\beta$}}\times\bf E']\\
{\bf E}' =  {\bf E}_\parallel + \gamma[{\bf E}_\perp + {\mbox{\boldmath $\beta$}}\times\bf B], 
& \,\,\,\,\,{\bf B}'=  {\bf B}_\parallel + \gamma[{\bf B}_\perp - {\mbox{\boldmath $\beta$}}\times\bf E]\;.
\end{eqnarray}
For a charge velocity  very small ($\beta<<1, \gamma\rightarrow 1$) the field transformations reduce to
\begin{eqnarray}
\nonumber 
{\bf E} = {\bf E}'- {\mbox{\boldmath $\beta$}}\times\bf B', & \,\,\,{\bf B} = {\bf B}'+ {\mbox{\boldmath $\beta$}}\times\bf E'\\
{\bf E}' = {\bf E}+ {\mbox{\boldmath $\beta$}}\times\bf B, & \,\,\,\,{\bf B}' = {\bf B}- {\mbox{\boldmath $\beta$}}\times\bf E\;.
\end{eqnarray}
Thus in our case of a Coulomb field ($\bf B'=0$) we get ${\bf E}={\bf E}'$ and ${\bf B}= {\mbox{\boldmath $\beta$}}\times\bf E'$, 
which in a spherical coordinate system ($R,\theta,\phi$), with ${\mbox{\boldmath $\beta$}}$ along $\theta=0$ 
(i.e., along the $z$-axis), can be written as,
\begin{eqnarray}
E_{R} & = & \frac{e}{R^{2}}  \\
B_{\phi} & = & \frac{e\beta \sin\theta}{R^{2}} \;,
\end{eqnarray}
where $R=R'$ (for $\beta<<1$, $\gamma\rightarrow 1$) is the distance between the field point and the ``present'' charge position, 
meaning the charge location determined simultaneous to the field.
Thus the electric field to a first order in $\beta$ is independent of velocity as well as angle and is simply the radial Coulomb 
field with respect to the ``present'' charge position. However there is an accompanying magnetic field proportional to $\beta\sin\theta$. 
\begin{figure}
\scalebox{0.65}{\includegraphics{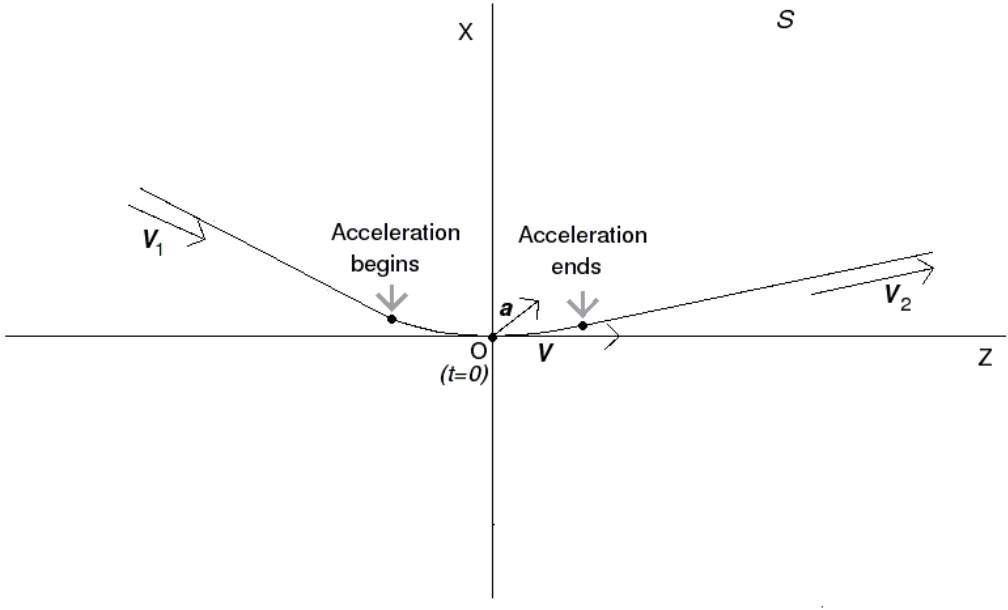}}
\caption{A schematic of the motion of the accelerated charge in the lab-frame ${\cal S}$. At $t=0$ the charge is moving with ${\bf V}$, 
the average of the velocities before and after the acceleration ${\bf a}$, which may not be parallel to ${\bf V}$.}
\end{figure}
\subsection{An accelerated charge}
Now we want to determine the fields of a charge that is subjected to acceleration. 
We assume that a charge initially moving with a velocity ${\bf V}_1$ (relativistic!) in the lab-frame (${\cal S}$) is subjected to 
an acceleration ${\bf a}$ (not necessarily parallel to ${\bf V}_1$) for a short duration $\Delta t$ so that its velocity changes by 
$\Delta {\bf V}$ (with $\Delta V<<c$) to become ${\bf V}_2$ (Fig.~1). We assume ${\bf a}=\Delta {\bf V}/\Delta t$ is a constant 
during the interval $\Delta t$ and that ${\bf V}_1$ and ${\bf V}_2$ are constant outside that interval. 
In limit we could take $\Delta t \rightarrow 0$. At mid-point the time interval 
of acceleration, say at $t=0$, the charge has a velocity ${\bf V}=({\bf V}_1+ {\bf V}_2)/2$. Without any loss of generality we 
could choose orientation of axes so that motion of the charge lies in the X-Z plane with ${\bf V}$ along the Z-axis..  
In general the net velocity change $\Delta {\bf V}$ and acceleration ${\bf a}$ 
could be resolved into components parallel ($\Delta {\bf V}_\parallel$, ${\bf a}_\parallel$) and perpendicular ($\Delta {\bf V}_\perp$, 
${\bf a}_\perp$) to ${\bf V}$.
\begin{figure}
\scalebox{0.65}{\includegraphics{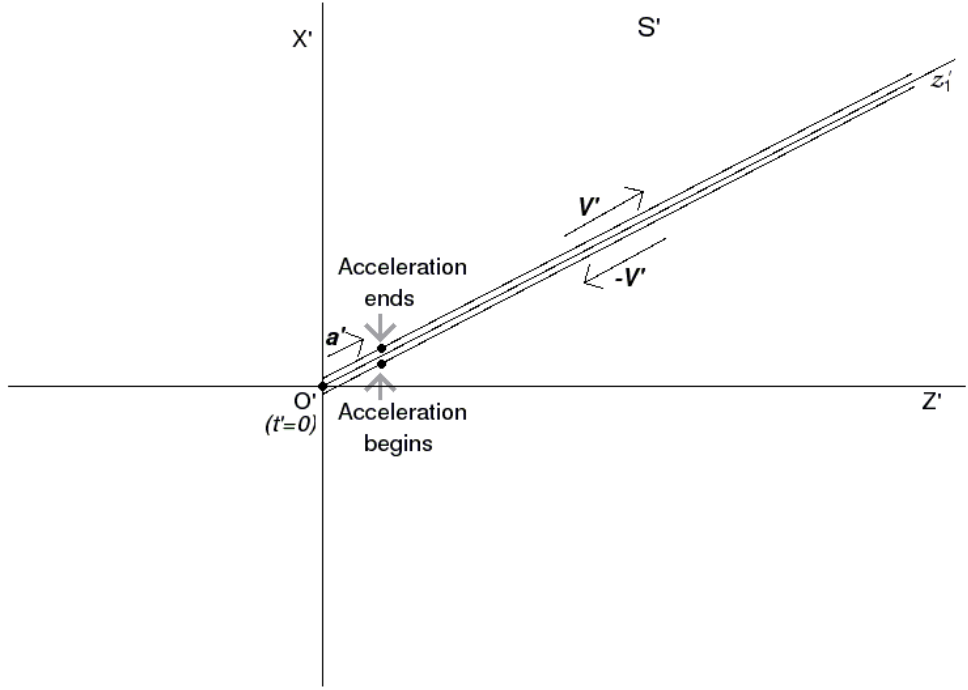}}
\caption{Motion of the accelerated charge as seen in the instantaneous rest-frame ${\cal S'}$. Two wings of the charge trajectory, 
which overlap before and after $t'=0$, have been displaced here slightly in a vertical direction for clarity.}
\end{figure}

We first determine EM fields in an inertial frame (${\cal S}'$) that is moving with a velocity ${\bf V}$ along the Z-axis 
with respect to the lab-frame (${\cal S}$), then frame ${\cal S}'$ will be an instantaneous rest frame of the charge at $t=0$. 
The charge in frame  ${\cal S}'$ will have the final velocity components 
${\bf V}'_{2\parallel}=\Delta {\bf V}_\parallel/[2(1-{\bf V}.{\bf V}_2/c^2)]
=\gamma^2\Delta {\bf V}_\parallel/2$ (for $\Delta V<<c$) and 
${\bf V}'_{2\perp}=\Delta {\bf V}_\perp/[2\gamma(1-{\bf V}.{\bf V}_2/c^2)]=\gamma\Delta {\bf V}_\perp/2$, 
where $\gamma=(1- V^2/c^2)^{-1/2}$ 
is the Lorentz factor. In the same way it can be seen that the initial velocity components in  ${\cal S}'$ will be 
${\bf V}'_{1\parallel}=-{\bf V}'_{2\parallel}$ and ${\bf V}'_{1\perp}=-{\bf V}'_{2\perp}$. We could then suppress the sub-scripts  
and write the initial velocity to be $-{\bf V}'$ and the final velocity to be ${\bf V}'$. Then acceleration, lasting for a time interval 
$\Delta t'=\Delta t/\gamma$, will have components ${\bf a}'_\parallel=2{\bf V}'_\parallel/\Delta t'=
\gamma^3{\bf a}_\parallel$ and ${\bf a}'_\perp=2{\bf V}'_\perp/\Delta t'=\gamma^2{\bf a}_\perp$. It is to be noted that the 
motion of the charge in the instantaneous rest frame (${\cal S}'$) will be essentially one-dimensional with ${\bf a}'\parallel{\bf V}'$, 
irrespective of the angle between  ${\bf a}$ and ${\bf V}$ in ${\cal S}$ (of course within our assumption $\Delta V<<c$). 
Thus in frame ${\cal S}'$ 
the charge initially moving with a velocity $-{\bf V}'=-{\mbox{\boldmath $\beta$}}'c$ will be subjected to an acceleration 
${\bf a}'=\dot{\mbox{\boldmath $\beta$}}'c$ for a brief time interval $\Delta t'$ so that it velocity finally becomes 
${\bf V}'={\bf a}'\:\Delta t'/2$ and henceforth it will continue to move with this uniform velocity ${\bf V}'$, retracing its earlier path 
(Fig.~2). For convenience in field calculations we can rotate the co-ordinate axes within ${\cal S}'$ 
so that the charge motion (along the $O'z'_1$ direction) is perceived to be along a horizontal axis. 
The frame ${\cal S}$ will then  be moving with a velocity $-{\bf V}$ along the inclined axis $O'Z'$ in ${\cal S}'$.
At the end we could undo the rotation or perhaps better still, write the field expressions in a co-ordinate independent language. 
We shall assume that the EM field expressions may depend only upon position, velocity and acceleration of the charge as specified 
at the retarded time and not on any higher temporal derivatives of its motion. This effectively implies 
that we need not worry about the temporal suddenness or sharpness of the applied acceleration. Further we make the reasonable 
assumption that the fields fall to zero at infinite distance from the charge.
\subsubsection{Electric field of the accelerated charge in the instantaneous rest-frame ${\cal S}'$}
\begin{figure}
\scalebox{0.65}{\includegraphics{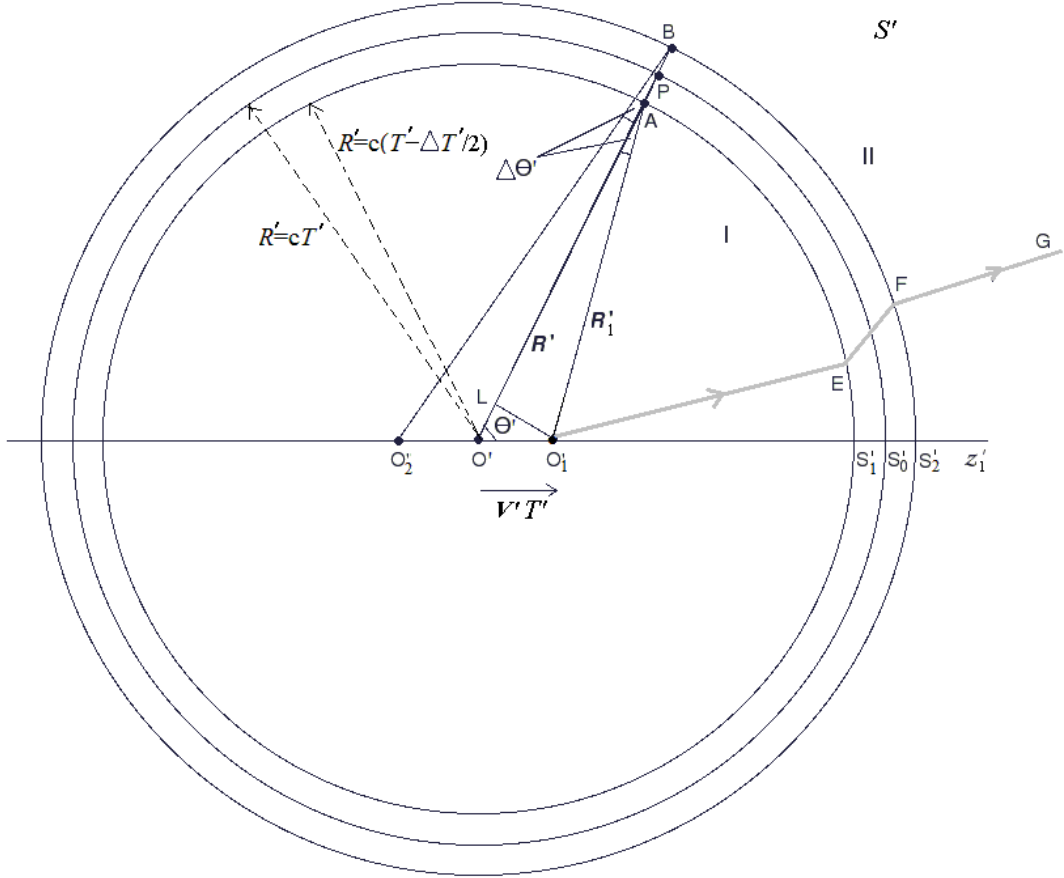}}
\caption{Calculation of the electric field of the accelerated charge. The faint grey line $O'_1EFG$ illustrates a representing 
electric field line as per Thomson's construction \cite{9} showing the `kink' in the region between $S'_1$ and $S'_2$ 
corresponding to the acceleration of the charge.}
\end{figure}
Let $t'=0$ be the time at which the charge was instantaneously at rest at say, $O'$. During the time of acceleration $\Delta t'$ 
the charge moves a distance $a'\:\Delta t'^2/8$ towards $O'$ and then 
back, and to a first order in $\Delta t'$ we can assume that during the time interval of acceleration, charge remains practically  
stationary at $O'$. Let us now consider the fields of the charge at a later time $T'>>\Delta t'$. 
At far off places with $R'>c(T'+\Delta t'/2)$ (Fig.~3, region II) the information could not have reached that the charge 
motion has undergone a change. Therefore the electric field there would continue to be that of the charge moving with its 
initial velocity $-{\bf V}'$, and will be thus directed in a radial direction (Eq.~4) from the extrapolated position of the charge 
($O'_2$ in Fig.~3) that it would have occupied at that moment had it not been subjected to an acceleration. 
But in the nearer regions with $R'<c(T'-\Delta t'/2)$ (region I) the field would have adjusted to the changed motion of 
the charge and will now be directed radially from point $O'_1$, the actual position of the charge at time $T'$. Only in the narrow 
shell of thickness $\Delta R'=c\Delta t'$ in the region between $S'_1$ and $S'_2$, which is causally 
connected to the time interval of acceleration at $O'$, the fields are yet to be determined.

We examine the electric field along $O'APB$, a radial direction from $O'$. In the absence of acceleration the field at $P$ 
would be just the Coulomb field, i.e., radial from $O'$ along $O'P$. Now we want to find out what field changes may take 
place when acceleration is imposed upon the charge. As we discussed above, the electric field at $A$ is along $O_1'A$ which is 
inclined at a small angle $\Delta\theta'$ with respect to $O'A$. 
To a first order in $\beta'$, $\Delta\theta'=O'_1L/R'_1=O'O'_1\sin \theta' /R'_1=V'T'\sin \theta' /R'_1=\beta'\sin \theta'$. 
Therefore the electric field vector at $A$ can be resolved into a radial component along $O'A$, plus a 
transverse component along $\hat{\theta'}$ which to a first order in $\beta'$ is $\Delta \theta' \,e/R'^2=e\beta' \sin\theta'/R^2$. 
Similarly at $B$ there is a transverse component $-e\beta' \sin\theta'/R'^2$.
Now as the shell sweeps past $P$, the transverse component of the electric field there will change 
from $-e\beta' \sin\theta'/R'^2$ at time ($T'-\Delta t'/2$)  to $e\beta' \sin\theta'/R'^2$ at ($T'+\Delta t'/2$), and irrespective of the 
duration $\Delta t'$ of acceleration it implies a zero value at $T'$, 
not surprising since at time $T'$ the field point $P$ has a causal relation to the charge at $O'$ when it had a zero velocity.

However, in addition to the temporal change at $P$, there is also a spatial change   
$-2e\beta' \sin\theta'/R'^2=-e\dot\beta' \Delta t'\sin\theta'/R'^2=-e\dot\beta' \Delta R'\sin\theta'/cR'^2$ 
in the transverse electric field component over a distance $\Delta R'$ from $A$ to $B$. The ratio 
$\Delta E'_{\theta'}/\Delta R'= -e\dot{\beta'}\sin\theta'/cR'^2$, a finite value, is independent of the width $\Delta R'$ of the shell, 
and in limit therefore we can write a gradient,
\begin{equation}
\frac{\partial E'_{\theta'}}{\partial R'}= \frac{-e}{c} \; \frac{\dot{\beta'} \sin\theta'}{R'^2}\;.
\end{equation}
There are discontinuities in the gradient at $B$ and $A$, basically due to the discontinuity in $\dot{\mbox{\boldmath $\beta'$}}$ 
at the retarded times $t'=-\Delta t'/2$ and $\Delta t'/2$ in the case considered here. But at $P$ it is ``well-behaved'' 
as the acceleration has a continuity at time $t'=0$.

Equation (6) at $P$ has a formal solution,  
\begin{equation}
\nonumber
E'_{\theta'}=   \frac{e \,\dot{\beta'}\,  \sin\theta'}{c\, R'} + \alpha\;.
\end{equation}
We can put the constant of integration $\alpha=0$, considering that the fields fall to zero as $R' \rightarrow \infty$ 
(in the absence of a discontinuity in $\dot{\mbox{\boldmath $\beta'$}}$). Thus whenever an acceleration is imposed on a 
stationary charge, it will result in an electric field gradient $\partial E'_{\theta'}/\partial R'= -e\dot{\beta'}\sin\theta'/cR'^2$ 
in the neighbourhood of the field point $P(R',\theta')$, in turn implying a finite acceleration dependent transverse field 
$E'_{\theta'}= {e \,\dot{\beta'}\,\sin\theta'}/{c\, R'}$. This will be in addition to the radial field component $E'_{R'} =e /R'^2$.
The total electric field at any point on spherical surface $S'_0$, causally connected to an accelerated charge when it had a 
zero velocity, can then be written in a co-ordinate independent manner as 
\begin{equation}
{{\bf E}'}=\frac {e\,{\bf n}'}{R'^2}+
\frac {e}{c} \,
\frac{{\bf n}'\times({\bf n}'\times\dot{\mbox{\boldmath $\beta'$}})}{R'} \;,
\end{equation}
where unit vector ${\bf n}'={\bf R}'/R'$.
\subsubsection{Magnetic field of the accelerated charge}
The magnetic field of the charge can be determined in a similar manner as electric field above. From Eq.~(5) there is a 
component of the magnetic field along $\hat\phi'$ that changes from $e \beta'  \sin\theta'/(R'^2)$ at ($R'-\Delta R'/2$) 
to $-e \beta'  \sin\theta'/(R'^2)$ at ($R'+\Delta R'/2)$. 
Therefore  at $P$, though the value of the above component may be zero, but  
there exists a finite gradient $\partial { B'}_{\phi'}/\partial R' =  -{e \,\dot{\beta'}\,\sin\theta'}/{cR'^2}$,
implying the presence of an acceleration dependent magnetic field,  
${ B'}_{\phi'}=  e \,\dot{\beta'}\,\sin\theta'/cR'$.

Thus on the spherical surface $S'_0$ we have a magnetic field
\begin{equation}
{\bf B'}= -\frac {e}{c\,R'}\,{\bf n'}\times \dot{\mbox{\boldmath $\beta'$}}={\bf n'}\times {\bf E'}  \;.
\end{equation}
It should be noted that the EM fields calculated above (Eqs.~7 and 8) are strictly true at points only on the spherical surface $S'_0$, i.e., 
corresponding to an accelerated charge that is instantaneously stationary at $O'$. These may not hold true unequivocally at points other 
than on surface $S'_0$ in the region between $S'_1$ and $S'_2$ when the charge had gained a finite velocity at the corresponding 
retarded time and the field expressions might need some velocity-dependent correction terms. The same are derived in the next section.
\section{Electromagnetic fields of an accelerated charge moving with a non-relativistic velocity}
Now we want to derive EM fields for an accelerated charge which has a finite but non-relativistic velocity 
${\mbox{\boldmath $\beta$}}={\bf V}/c$ in an inertial frame say, ${\cal S}$. We 
assume that at $t=0$ the charge is at the origin $O$, moving with a velocity ${\mbox{\boldmath $\beta$}}$ and 
acceleration $\dot{\mbox{\boldmath $\beta$}}$ ($\dot{\mbox{\boldmath $\beta$}}$ not necessarily parallel to 
${\mbox{\boldmath $\beta$}}$). We want to calculate the EM fields at point $D$, along radius vector ${\bf R}$ from $O$, 
at time $T=R/c$ (Fig.~4). 
Let ${\cal S}'$ be an inertial frame, moving with a velocity ${\mbox{\boldmath $\beta$}}$ with respect to ${\cal S}$, 
with their origins $O'$ and $O$ coinciding at $t'=t=0$. In frame ${\cal S}'$, $O'$ is the retarded time position of the 
instantaneously stationary charge (the charge will not forever remain stationary at $O'$ because of its acceleration), 
corresponding to the field point $D$, at radius vector ${\bf R}'$ from $O'$.

From a non-relativistic Lorentz transformation ($\beta<<1, \gamma\rightarrow 1$) between $\cal S$ and $\cal S'$ we have,
\begin{eqnarray}
{\bf R}'& = & {\bf R}-{\bf V}T= {\bf R}-{\mbox{\boldmath $\beta$}}R=R({\bf n}-{\mbox{\boldmath $\beta$}})\\
T' & = & T-{\bf V}.{\bf R}/c^2 = T(1-{\mbox{\boldmath $\beta$}}.{\bf n})
\end{eqnarray}
It should be noted that a non-relativistic Lorentz transformation does not necessarily imply a 
Galilean transformation. In particular $T'=T(1-{\mbox{\boldmath $\beta$}}.{\bf n})$, a non-Galilean feature, 
is a must if the fields have to propagate 
with the same speed $c$ in both reference frames. From Eqs. (9) and (10) we get,
\begin{equation}
R'=cT'=R(1-{\mbox{\boldmath $\beta$}}.{\bf n})
\end{equation}
\begin{equation}
{\bf n}'={\bf R}'/R'=({\bf n}-{\mbox{\boldmath $\beta$}})/(1-{\mbox{\boldmath $\beta$}}.{\bf n}).
\end{equation}
\begin{figure}
\scalebox{0.6}{\includegraphics{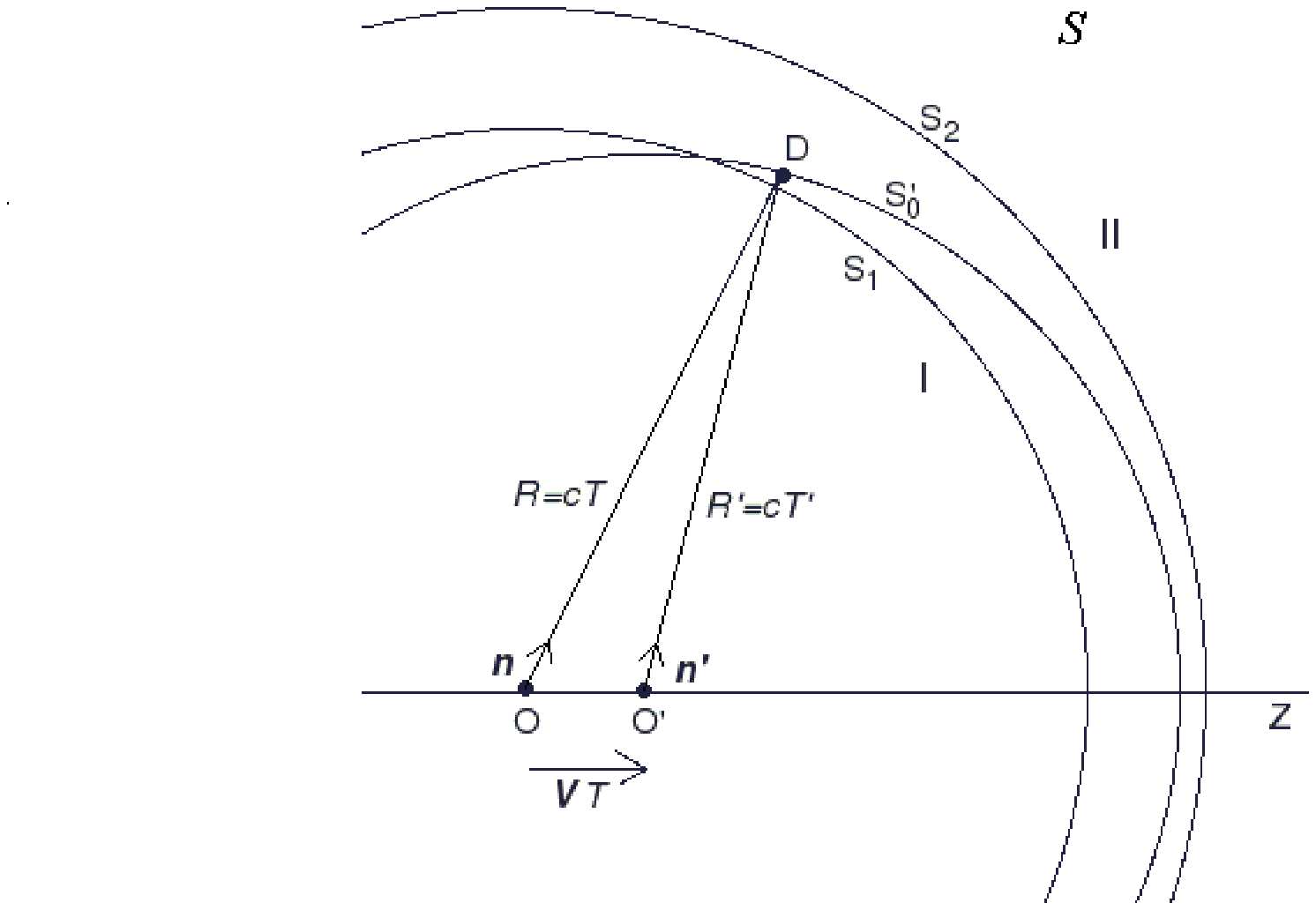}}
\caption{Calculation of the Electric field of the accelerated charge moving with a non-relativistic velocity}
\end{figure}

Electromagnetic fields at point $D$ on surface $S'_0$ in instantaneous rest-frame ${\cal S}'$ are given by Eqs.~(7) and (8). 
Using Eq.~(3) and noting that for a non-relativistic case $\dot{\mbox{\boldmath $\beta'$}}=\dot{\mbox{\boldmath $\beta$}}$, we 
can transform the electric field for ${\cal S}$ to get,
\begin{equation}
\nonumber
{{\bf E}}=\frac {e\,{\bf n}'}{R'^2}+ \frac {e}{c} \,\left[
\frac{{\bf n}'\times({\bf n}'\times\dot{\mbox{\boldmath $\beta$}})}{R'}+
\frac{{\mbox{\boldmath $\beta$}}\times({\bf n}'\times\dot{\mbox{\boldmath $\beta$}})}{R'}\right] 
\end{equation}
or
\begin{equation}
{{\bf E}}=\frac {e\,{\bf n}'}{R'^2}+ \frac {e}{c} \,
\frac{({\bf n}'+{\mbox{\boldmath $\beta$}})\times({\bf n}'\times\dot{\mbox{\boldmath $\beta$}})}{R'} \;.
\end{equation}
In this expression ${\bf n}'$ and $R'$, yet to be transformed in terms of quantities in ${\cal S}$, are specified still with 
respect to ${\cal S}'$. Now in frame ${\cal S}$, the
point $O'$ will be at ${\bf V}T={\mbox{\boldmath $\beta$}}R$ at time $T$, and the position vector of the field point with respect to $O'$ in 
${\cal S}$, (${\bf R}-{\bf V}T$, Fig.~4), is the same as ${\bf R}'$ in ${\cal S}'$ (Eq.~(9)). 
Thus ${\bf n}'$ and $R'$, which are the unit vector and the distance specified 
with respect to the position of the charge at the retarded time in ${\cal S}'$, are also the unit vector and the distance in ${\cal S}$ 
but specified with respect to the ``present'' positions ${\bf V}T$ of the charge, determined at time $T$ from the 
velocity value at the retarded time. The geometry of $O$, $O'$ and $D$ in Fig.~4 is similar to that 
of $O'$, $O'_1$ and $A$ in Fig.~3.  

Thus the electric field at any point in the region enclosed between surfaces $S_1$ and $S_2$ (Fig.~4) can be calculated 
directly from Eq.~(13), 
with ${\bf n}'$ and $R'$ specified with respect to the ``present'' position of the charge, calculated using the corresponding 
retarded time value of velocity. In fact Eq.~(13) gives electric field for every point in ${\cal S}$ (including regions I and II, Fig.~4), 
with the understanding that corresponding retarded time values of velocity and acceleration (if any) are to be used and 
${\bf n}'$ and $R'$ are to be specified with respect to the ``present'' position ${\bf V}T={\mbox{\boldmath $\beta$}}R$ of the charge 
calculated using the velocity ${\mbox{\boldmath $\beta$}}$ at the corresponding retarded time. Equation~(13) is thus a generalization of 
Eq.~(4) when acceleration is present.

From Eqs.~(11) and (12) we can substitute for ${\bf n}'$ and $R'$ in terms of the retarded time 
quantities ${\bf n}$ and $R$, where we also note that to a first order in ${\mbox{\boldmath $\beta$}}$,
${\bf n}'+{\mbox{\boldmath $\beta$}}={\bf n}/(1-{\mbox{\boldmath $\beta$}}.{\bf n})$. Then we get
\begin{equation}
{{\bf E}}=e\,\left[\frac {({\bf n}-{\mbox{\boldmath $\beta$}})}{R^2(1-{\mbox{\boldmath $\beta$}}.{\bf n})^3}\right]+ 
\frac {e}{c} \,\left[\frac{{\bf n}\times\{{({\bf n}-{\mbox{\boldmath $\beta$}})\times\dot{\mbox{\boldmath $\beta$}}\}}}
{R(1-{\mbox{\boldmath $\beta$}}.{\bf n})^3}\right] \;.
\end{equation}
All quantities on the right hand side are to be evaluated at the retarded time.
This expression supposed to be accurate to only first order in velocity, surprisingly gives all the terms correctly, especially 
for the acceleration fields (only missing a factor $1/\gamma^2$ in the velocity fields), when compared with the 
more general expression for EM fields (cf. ref.~\onlinecite{1,2,3,12}; also see Eq.~(27), Section IV here). The derivation shows  
the genesis of various terms in the field expressions, in particular that of the somewhat intriguing factor 
${\bf n}\times\{{({\bf n}-{\mbox{\boldmath $\beta$}})\times\dot{\mbox{\boldmath $\beta$}}\}}$ in the acceleration fields. 

The magnetic field can be expressed more easily in terms of the electric field from Eqs.~(3) and (8),
\begin{eqnarray}
\nonumber
{\bf B}& = & {\bf B}'+ {\mbox{\boldmath $\beta$}}\times\bf E'
 =  ({\bf n}'+{\mbox{\boldmath $\beta$}})\times\bf E'\\
& = & {\bf n}\times{\bf E}' /(1-{\mbox{\boldmath $\beta$}}.{\bf n})
\end{eqnarray}
Substituting for $\bf E'$ from Eq.~(3), we can rewrite the above equation as, 
\begin{equation}
\nonumber
{\bf B}(1-{\bf n} \;.\; {\mbox{\boldmath $\beta$}}) 
 =  {\bf n}\times {\bf E}+{\bf n}\times({\mbox{\boldmath $\beta$}}\times\bf B)
\end{equation}
\begin{equation}
\nonumber
{\bf B}= {\bf n}\times {\bf E}+{\mbox{\boldmath $\beta$}}({\bf n}\;.\; {\bf B})
\end{equation}
From Eq.~(15), ${\bf n}\;.\;{\bf B}=0$, giving us,
\begin{equation}
{\bf B}= {\bf n}\times {\bf E}
\end{equation}
\section{General expressions for electromagnetic fields of the accelerated charge}
Equations (14) and (16) give expressions for EM fields of an accelerated charge moving with a non-relativistic 
velocity. Our aim now is to derive expressions for EM fields of an accelerated charge moving with an arbitrary velocity 
${\mbox{\boldmath $\beta$}}={\bf V}/c$ in the lab-frame ${\cal S}$ and find changes, if any, required in the 
field expressions. Let ${\cal S}'$ be the inertial frame, moving with 
velocity ${\mbox{\boldmath $\beta$}}$ with respect to ${\cal S}$ and be thus an instantaneous rest frame of the charge. 
We resolve all vectors into parallel and perpendicular components by taking the former along the 
direction of motion, i.e., along ${\mbox{\boldmath $\beta$}}$. The distance vectors ${\bf R}$ and ${\bf R}'$ connect the field points 
to the time retarded positions of the charge while ${\bf n}={\bf R}/R$ and ${\bf n}'={\bf R}'/R'$ are the corresponding 
unit vectors directed from the time retarded positions towards the field point.

Then we can write ${\bf n}'={\bf n}'_\parallel+{\bf n}'_\perp$ with 
\begin{equation}
{\bf n}'_\parallel=\delta\,\gamma\,  ({\bf n}_\parallel - {\mbox{\boldmath $\beta$}})\,, \,\,\,\,\,\,\,\,{\bf n}'_\perp =\delta\, {\bf n}_\perp
\end{equation}
where $\delta = \gamma\,(1+{\bf n}' \;.\; {\mbox{\boldmath $\beta$}})=1/[{\gamma\,(1-{\bf n} \;.\; {\mbox{\boldmath $\beta$}})}]$
is the Doppler factor.\cite{1,13,14} 
Further,
\begin{equation}
R' = R \gamma\,(1-{{\bf n}} \;.\; {\mbox{\boldmath $\beta$}})= \delta^{-1}R
\end{equation}
From the above relations we also have,
\begin{equation}
\gamma\,({\bf n}'_\parallel + {\mbox{\boldmath $\beta$}})=\delta\, {\bf n}_\parallel \,, \,\,\,\,\,\,\,\,
1+{\bf n}' \;.\; {\mbox{\boldmath $\beta$}}=\delta^2\,(1-{\bf n} \;.\; {\mbox{\boldmath $\beta$}}).
\end{equation}
For the acceleration vector  
$\dot{\mbox{\boldmath $\beta'$}}=\dot{\mbox{\boldmath $\beta'$}}_\parallel
+\dot{\mbox{\boldmath $\beta'$}}_\perp$ we have the transformations,\cite{1,13,14} 
\begin{equation} 
\dot{\mbox{\boldmath $\beta'$}}_\parallel = \gamma^3\,\dot{\mbox{\boldmath $\beta$}}_\parallel, \;\;\;\;\;\; 
\dot{\mbox{\boldmath $\beta'$}}_\perp = \gamma^2\,\dot{\mbox{\boldmath $\beta$}}_\perp.
\end{equation}
Then we have the relation,
\begin{eqnarray}
\nonumber
({\bf n}' + {\mbox{\boldmath $\beta$}}) \;.\; \dot{\mbox{\boldmath $\beta'$}} & 
= & ({\bf n}'_\parallel + {\mbox{\boldmath $\beta$}}) \;.\; \dot{\mbox{\boldmath $\beta'$}}_\parallel 
+{\bf n}'_\perp \;.\; \dot{\mbox{\boldmath $\beta'$}}_\perp \\ 
 & = & \delta\, \gamma^2\,({\bf n}_\parallel \;.\; \dot{\mbox{\boldmath $\beta$}}_\parallel
+{\bf n}_\perp \;.\; \dot{\mbox{\boldmath $\beta$}}_\perp) =
\delta\, \gamma^2\,{\bf n} \;.\; \dot{\mbox{\boldmath $\beta$}}\;.
\end{eqnarray}
Another vector relation that we will need is,
\begin{eqnarray}
%\nonumber
{\mbox{\boldmath $\beta$}}\times({\bf n}'\times \dot{\mbox{\boldmath $\beta'$}}) & = &
%{\bf n}'({\mbox{\boldmath $\beta$}}\;.\; \dot{\mbox{\boldmath $\beta'$}})-
%\dot{\mbox{\boldmath $\beta'$}}({\bf n}' \;.\; {\mbox{\boldmath $\beta$}})\\ & = &
{\bf n}'_\perp({\mbox{\boldmath $\beta$}}\;.\; \dot{\mbox{\boldmath $\beta'$}})-
\dot{\mbox{\boldmath $\beta'_\perp$}}({\bf n}' \;.\; {\mbox{\boldmath $\beta$}})
\end{eqnarray}
which follows from,
\begin{equation}
{\bf n}'_\parallel({\mbox{\boldmath $\beta$}}\;.\; \dot{\mbox{\boldmath $\beta'$}})=
\dot{\mbox{\boldmath $\beta'_\parallel$}}({\bf n}' \;.\; {\mbox{\boldmath $\beta$}})=
n'_\parallel\, \dot{\beta'_\parallel}\,{\mbox{\boldmath $\beta$}}\;.
\end{equation}
\subsection{Electric field}
From the Lorentz transformation of EM fields (Eq.~(2)), we can write 
\begin{equation}
{\bf E}_\parallel={\bf E}'_\parallel\;,
\;\;\;\;\;\; {\bf E}_\perp=\gamma[{\bf E}'_\perp - {\mbox{\boldmath $\beta$}}\times\bf B'].
\end{equation}
Using Eqs.~(7) and (23), we can write
\begin{equation}
\nonumber
{{\bf E}_\parallel}=\frac {e\,{\bf n}'_\parallel}{R'^2}+\frac {e}{c\,R'} 
[{\bf n}'_\parallel\{({\bf n}' + {\mbox{\boldmath $\beta$}})\;.\; \dot{\mbox{\boldmath $\beta'$}}\}-
\dot{\mbox{\boldmath $\beta'_\parallel$}}(1+{\bf n}' \;.\; {\mbox{\boldmath $\beta$}})] \;.
\end{equation}
Substituting from Eqs.~(17-21), we get
\begin{equation}
{\bf E}_\parallel=\frac {e\,\gamma\, \delta^3}{R^2} 
({\bf n}_\parallel-{\mbox{\boldmath $\beta$}})+\frac {e\,\gamma^3\, \delta^3}{c\,R} 
[({\bf n}_\parallel-{\mbox{\boldmath $\beta$}})({\bf n}\;.\; \dot{\mbox{\boldmath $\beta$}})-
\dot{\mbox{\boldmath $\beta_\parallel$}}(1-{\bf n} \;.\; {\mbox{\boldmath $\beta$}})] \;.
\end{equation}
In the same way from Eqs.~(7), (8) and (24) we have,
\begin{equation}
\nonumber
{\bf E}_\perp=\frac {e\,\gamma\,{\bf n}'_\perp}{R'^2}+
\frac {e\,\gamma}{c\,R'} 
\{{\bf n}'_\perp({\bf n}'\;.\; \dot{\mbox{\boldmath $\beta'$}})-
\dot{\mbox{\boldmath $\beta'_\perp$}}\} \;
 + \frac {e\,\gamma}{c\,R'}{\mbox{\boldmath $\beta$}}\times({\bf n}'\times \dot{\mbox{\boldmath $\beta'$}}),
\end{equation}
Using Eq.~(22), we can write it as,
\begin{equation}
\nonumber
{\bf E}_\perp=\frac {e\,\gamma\,{\bf n}'_\perp}{R'^2}+
\frac {e\,\gamma}{c\,R'} 
[{\bf n}'_\perp\{({\bf n}' + {\mbox{\boldmath $\beta$}})\;.\; \dot{\mbox{\boldmath $\beta'$}}\}-
\dot{\mbox{\boldmath $\beta'_\perp$}}(1+{\bf n}' \;.\; {\mbox{\boldmath $\beta$}})] \;.
\end{equation}
Substituting from Eqs.~(17-21), we get
\begin{equation}
{\bf E}_\perp=\frac {e\,\gamma\, \delta^3}{R^2} 
{\bf n}_\perp+\frac {e\,\gamma^3\, \delta^3}{c\,R} 
[{\bf n}_\perp({\bf n}\;.\; \dot{\mbox{\boldmath $\beta$}})-
\dot{\mbox{\boldmath $\beta_\perp$}}(1-{\bf n} \;.\; {\mbox{\boldmath $\beta$}})] \;.
\end{equation}
Adding Eqs.~(25) and (26), we get,
\begin{equation}
\nonumber
{{\bf E}}=\frac {e\,\gamma\, \delta^3}{R^2} 
({\bf n}-{\mbox{\boldmath $\beta$}})+\frac {e\,\gamma^3\, \delta^3}{c\,R} 
[({\bf n}-{\mbox{\boldmath $\beta$}})({\bf n}\;.\; \dot{\mbox{\boldmath $\beta$}})-
\dot{\mbox{\boldmath $\beta$}}(1-{\bf n} \;.\; {\mbox{\boldmath $\beta$}})] 
\end{equation}
which can be expressed in a more familiar form (cf. ref.~\onlinecite{1,2,3,12})
\begin{equation}
{\bf E}=e \left[
\frac{({\bf n}-\mbox{\boldmath $\beta$})
}{R^2\,\gamma^2(1-{{\bf n}}\;.\;\mbox{\boldmath $\beta$})^{3}}\right]
+\frac {e}{c} \,\left[
 \frac{{{\bf n}}\times\{({{\bf n}}-\mbox{\boldmath $\beta$})\times
\dot{\mbox{\boldmath $\beta$}}\}}{R\,(1-{{\bf n}}\;.\;\mbox{\boldmath $\beta$})^{3}} \right]\;.
\end{equation}
All quantities on the right hand side are to be evaluated at the retarded time. The expression differs from its 
non-relativistic counter-part (Eq.~(14)) in only an extra term $1/\gamma^2$ in velocity fields (first term in square brackets 
on the right hand side).

\subsection{Magnetic field}
One can explicitly calculate ${\bf B}$ following a similar procedure as above. However it is much more 
simpler to express ${\bf B}$ in terms of ${\bf E}$. Using Eqs.~(2), (8), (17) and (19) we can write,
\begin{eqnarray}
\nonumber
{\bf B}& = & {\bf B}'_\parallel + \gamma[{\bf B}'_\perp + {\mbox{\boldmath $\beta$}}\times\bf E']\\
\nonumber
& = & {\bf n}'_\perp\times {\bf E}'_\perp+\gamma[{\bf n}'_\perp\times{\bf E}'_\parallel
+{\bf n}'_\parallel\times {\bf E}'_\perp+{\mbox{\boldmath $\beta$}}\times{\bf E'}_\perp]\\
\nonumber
& = & [{\bf n}'_\perp+\gamma({\bf n}'_\parallel+{\mbox{\boldmath $\beta$}})] \times {\bf E}'_\perp+\gamma\,{\bf n}'_\perp
\times{\bf E}'_\parallel\\
\nonumber
& = & \gamma\, \delta\, [({\bf n}_\perp+{\bf n}_\parallel)\times\{{\bf E}_\perp + 
{\mbox{\boldmath $\beta$}}\times\bf B\}+{\bf n}_\perp\times {\bf E}_\parallel]\\
& = & \gamma\, \delta\,[{\bf n}\times {\bf E}+{\bf n}\times({\mbox{\boldmath $\beta$}}\times\bf B)]
\end{eqnarray}
We can rewrite Eq.~(28) as,
\begin{equation}
\nonumber
{\bf B}(1-{{\bf n}}\;.\;\mbox{\boldmath $\beta$})= {\bf n}\times {\bf E}+{\bf n}\times({\mbox{\boldmath $\beta$}}\times\bf B)
\end{equation}
\begin{equation}
\nonumber
{\bf B}={\bf n}\times {\bf E}+{\mbox{\boldmath $\beta$}}({\bf n}\;.\; {\bf B})
\end{equation}
From Eq.~(28), ${\bf n}\;.\;{\bf B}=0$. Thus we get the required relation,
\begin{equation}
\nonumber
{\bf B}= {\bf n}\times {\bf E}
\end{equation}
which is identical to the non-relativistic expression (Eq.~(16)).
\section{Conclusions}
We have derived the EM fields of a charge with an arbitrary motion. The expressions for 
both the electric and magnetic fields were first
derived for a charge in its instantaneous rest frame using a physical picture. 
Thereafter we calculate the EM fields of an accelerated charge having a non-relativistic motion. The expressions for these 
fields, accurate to first order in velocity $\beta$, when expressed with respect to the time retarded position of the charge, 
surprisingly yield all terms exactly for the acceleration fields. A Lorentz transformation from the instantaneous rest frame, 
using relativistic Doppler factors, then 
led us to the standard expressions for the electromagnetic fields of an arbitrarily moving charge without using 
 Li\'{e}nard-Wiechert potentials and without resorting to any differentiation that need to be evaluated at the retarded time. 
The derived expressions of course agree with those derived from Li\'{e}nard-Wiechert potentials.

\end{document}